\documentstyle[12pt]{article}

\begin{titlepage}
\title{
{
\begin{flushright}
\vspace{-10ex}
{\small IHEP 2001--25~~~~~~~~~~~}\\[-1ex]
{\small KIAS--P01022~~~~~~~~~~~}\\[-1ex]
{\small BROWN--HET--1262~~}\\[-1ex]
{\small DPDG/01/01~~~~~~~~~~~}\\
\end{flushright}
}
\vspace{1ex}
{\bf A compilation of total cross section data}\\
{\bf on $e^+{}e^-\to {hadrons}$ and pQCD tests}\\
}

\author{O.\ V.\ Zenin, V.\ V.\ Ezhela,
S.\ B.\ Lugovsky\\
{\small COMPAS Group, IHEP, Protvino, Russia}\\[5mm]
M.\ R. \ Whalley\\
{\small HEPDATA Group, Durham University, Durham, UK}\\[5mm]
K.\ Kang$^{\rm a,b}$ and S.\ K.\ Kang$^{\rm a}$ \\
{\small $^{\rm a}$Korea Institute for Advanced Study, 
$^{\rm b}$Brown University}\\
}
 
\date{}
     
\end{titlepage}

\begin{document}
\maketitle

\abstract{
\noindent
All available data on the total cross sections and $R$-ratio of
$e^+{}e^-\to {hadrons}$ are compiled from the  PPDS (DataGuide),
PPDS (ReacData) and HEPDATA (Reaction) databases and transformed to a
compilation of data on the $e^+{}e^-\to quark~\overline{quark} \to {hadrons}$
in the continuum region
which can be used in tests of the parton model and pQCD calculations. This evaluated data
compilation is made available in PPDS system and is accessible through the Web.
It is shown that current predictions from the parton model and pQCD are 
well supported by this world ``continuum" data compilation which can
then be used in future refinements of the $\alpha_s(Q^2)$ as well as $\alpha_{QED}(Q^2)$
evolution forms. 
}

\section*{Motivation}
Despite the tremendous efforts of experimentalists and phenomenologists devoted
to the investigation of hadron production in $e^+ e^-$ collisions
there is no real integrated view on the experimental situation even for one of
the main observables, the total cross section for the reaction $e^+{}e^-\to hadrons$ and the ratio
$$ R ={{ \sigma(e^+ e^- \to hadrons)} \over 
{\sigma(e^+ e^- \to \mu^+ \mu^-)_{QED}}}$$

\noindent
Some data are published only as cross sections, $\sigma$, and others only as $R$.
The situation is more complicated as QED radiative effects to the data
have been corrected for in different ways, and to different degrees, in the various
data sets.
For example, in constructing $R$ experimentalists have used
two forms of theoretical calculations, one with $\alpha_{QED}(0)$, or another
taking into account the then currently accepted form of evolution of $\alpha_{QED}(Q^2)$.
Furthermore, some data were never published in numerical form and have been read by
compilers from graphs. The current experimental situation (March 2001) is 
summarized in Figures 3 and 4.

This aim of this work is firstly to prepare a comprehenesive compilation of
$\sigma$ and $R$ transformed, wherever possible,  to a unique, and meaningful, style of
implementing the pure QED radiative corrections. The second aim is to construct a
procedure to extract ``continuum data sets" where the direct comparison
of the parton model with different variants of the pQCD corrections will be 
conclusive. In selecting data for the compilation we use expert assessments
from recent reviews and works dealing with precision estimation of running 
$\alpha_{QED}(Q^2)$ and $\alpha_{S}(Q^2)$ (see 
~\cite{Schopper:1992ut},
~\cite{Eidelman:1995ny},
~\cite{Burkhardt:1995tt},
~\cite{Swartz:1996hc},
~\cite{Alemany:1998tn}, 
~\cite{Davier:1998si},
~\cite{Bethke:2000ai},
~\cite{Hinchliffe:2000yq},
~\cite{Zhao:2000xc}).


\section{Data selection and normalizations}

\subsection{General overview}
\noindent
 The data, which have been published in \cite{Behrend:1987md}-\cite{Dolinsky:1991vq},
were extracted in  numerical form from the PPDS RD \cite{PPDS} database and from the
Reaction database of HEPDATA system \cite{hepdata}. The following criteria were
then applied to define suitable data sets:

\begin{itemize}
\item All preliminary data are excluded;
\item Data obtained under incomplete kinematical conditions,
and not extrapolated to the complete kinematic region, are excluded;
\item Data in the form of dense energy scans are omitted if the authors published
in addition the data averaged over wider energy bins.
\end{itemize}

The data sets remaining after the above exclusion criteria have been applied, are
subdivided into the following six cathegories:\\

\boldmath
{\large\bf $\sigma_1^{sd}$} 
\unboldmath
(Refs.~\cite{Berger:1979uv} -- \cite{Miyabayashi:1995ej}).
Cross section data corrected by their authors
for the contributions of two-photon exchange diagrams,
the initial state bremsstrahlung and for the initial state vertex loops,
but not corrected for the vacuum polarization contribution to
the running $\alpha_{QED}$. These  ``{\bf semi-dressed}" cross sections
correspond to
the contribution of the diagram shown on Fig.~1.
The full $\gamma/Z$ propagator, taking into account
all vacuum polarization effects,
is denoted on the figure by the bold waved line.\\


\boldmath
{\large\bf $\sigma_2^{sd}$} 
\unboldmath
(Refs.~\cite{Cosme:1974xx} -- \cite{Augustin:1975yq}). 
Cross section data radiatively corrected by their authors according to the procedure
of Bonneau and Martin \cite{Bonneau:1971mk}.
This procedure included radiative corrections for the
initial state radiation, electronic vertex correction, and
the correction for the electronic loop in the photon propagator.
Thus it partially took into account 
the vacuum polarization effects also.
To obtain from these cross sections the ``{\bf semi-dressed}" ones
(corresponding to the diagram on Fig.~1) we rescale
\boldmath
$\sigma_2^{sd}$
\unboldmath 
by a factor
$1/(1-\Delta\alpha_{QED}^e{}(s))^2$.
%
We suggest that all the cross sections published before 1978
were radiatively corrected according to this procedure.\\
\begin{picture}(200,100)(0,0)

\put(25,50){\oval(50,50)[r]}
\put(50,50){\circle*{5}}
\put(25,75){\vector(4,0){10}}
\put(30,25){\vector(-4,0){1}}

\put(25,80){$e^-$}
\put(25,10){$e^+$}

\multiput(49.5,50)(10,0){10}{\oval(5,5)[t]}
\multiput(54.5,50)(10,0){10}{\oval(5,5)[b]}
\multiput(49.5,49)(10,0){10}{\oval(5,5)[t]}
\multiput(54.5,49)(10,0){10}{\oval(5,5)[b]}
\multiput(49.5,48)(10,0){10}{\oval(5,5)[t]}
\multiput(54.5,48)(10,0){10}{\oval(5,5)[b]}

\multiput(49.75,50)(10,0){10}{\oval(5,5)[t]}
\multiput(54.75,50)(10,0){10}{\oval(5,5)[b]}

\multiput(50,50)(10,0){10}{\oval(5,5)[t]}
\multiput(55,50)(10,0){10}{\oval(5,5)[b]}

\multiput(50,49)(10,0){10}{\oval(5,5)[t]}
\multiput(55,49)(10,0){10}{\oval(5,5)[b]}

\multiput(50.25,48)(10,0){10}{\oval(5,5)[t]}
\multiput(55.25,48)(10,0){10}{\oval(5,5)[b]}

\multiput(50.5,48)(10,0){10}{\oval(5,5)[t]}
\multiput(55.5,48)(10,0){10}{\oval(5,5)[b]}

\put(200,50){\oval(100,70)[l]}
\put(200,50){\oval(100,50)[l]}
\put(200,50){\oval(100,30)[l]}
\put(200,50){\oval(100,10)[l]}

\put(150,50){\circle*{20}}

\put(160,0){$hadrons$}


\put(90,60){$\gamma$,$Z$}
\put(80,-10){Fig.~1}

\end{picture}
\begin{picture}(200,100)(-20,0)

\put(25,50){\oval(50,50)[r]}
\put(50,50){\circle*{5}}
\put(25,75){\vector(4,0){10}}
\put(30,25){\vector(-4,0){1}}

\put(25,80){$e^-$}
\put(25,10){$e^+$}

\multiput(50,50)(10,0){10}{\oval(5,5)[t]}
\multiput(55,50)(10,0){10}{\oval(5,5)[b]}

\put(200,50){\oval(100,70)[l]}
\put(200,50){\oval(100,50)[l]}
\put(200,50){\oval(100,30)[l]}
\put(200,50){\oval(100,10)[l]}

\put(150,50){\circle*{20}}

\put(160,0){$hadrons$}


\put(90,60){$\gamma$,$Z$}
\put(80,-10){Fig.~2}

\end{picture}
\\
\\
\\

\boldmath
{\large $R_3^{bare}$}
\unboldmath
(Ref.~\cite{Brandelik:1978ei}).~Data on $R$
obtained from measurements of the ``{\bf semi-dressed}" 
\boldmath
($\sigma^{sd}_2$)
\unboldmath
cross-section divided, by their authors, by the point-like muonic
cross section with fixed $\alpha_{QED} = \alpha_{QED}(0)$
We rescale these data  by the factor
$(\alpha_{QED}(0)/\alpha_{QED}(s))^2/(1-\Delta\alpha^e_{QED}(s))^2$,
to obtain the ``{\bf bare}" $R$-parameter described by the diagram shown
on the Fig.~2. The tree-level $\gamma/Z$ propagator is denoted here
by the waved line.\\ 

\boldmath
{\large $R_4^{bare}$} 
\unboldmath
(Refs.~\cite{Behrend:1987md} -- \cite{Bai:2001ct}).
Data on $R$  obtained from measurements of the ``{\bf semi-dressed}"
\boldmath
($\sigma^{sd}_1$) 
\unboldmath
cross section divided, by their authors,
by the point-like muonic cross section,
but with the running $\alpha_{QED}(s)$ thus obtaining the ``{\bf bare}" 
$R$-ratio, which corresponds to the Fig.~2 diagram again.\\

\boldmath
{\large $\sigma_5^{sd}$} 
\unboldmath
(Refs.~\cite{Acton:1993yc} -- \cite{Acciarri:2000rw}).
~LEP I cross section data at the $Z$ peak not corrected for initial
state radiation and
electronic QED vertex loops, and LEP II -- III cross section data at
$\sqrt{s} > 130$ GeV
with a cut \mbox{$s'/s = 1 - (\sqrt{s}/2)(1 - 0.7225)$}.
One can interprete $\sqrt{s'}$ as an effective mass
of the propagator after the initial state radiation has reduced
the $e^+{}e^-$ pair center-of-mass energy.
Indeed, the definition of $s'$ depends on the assumptions
about the initial and final state radiation interference.
These data were rescaled to the ``{\bf semi-dressed}" cross section,
\boldmath
$\sigma_1^{sd}$
\unboldmath
,  defined by the
contribution of the Fig.~1 diagram as follows.
First, the theoretical cross sections\footnote{Values for
$\sigma^{th}_{cut}$ were calculated using ZFITTER 
subroutine ZUTHSM with argument settings according to the cuts applied
in the LEP I-II-III measurements.
Values for $\sigma^{th}_{born}$ were calculated by the same subroutine but 
with the special flag switching it to calculate the IBA cross sections}
 ~$\sigma^{th}_{cut}$  and 
$\sigma^{th}_{Born}$ were calculated using the ZFITTER 6.30
package\footnote{ZFITTER 6.30 Fortran code is available at 
http://www.ifh.de/$^{\sim}$riemann/Zfitter/zfitr6\underline{~}30.uu~.}
~\cite{Bardin:2001yd}
assuming the recommended by PDG~\cite{Groom:2000in} values of the standard model
parameters.
$\sigma^{th}_{cut}$ means here the cross section measured
with cuts applied in the particular experiment and calculated 
by ZFITTER.
$\sigma^{th}_{Born}$ denotes theoretical cross section,
calculated by ZFITTER 
in the Improved Born Approximation (IBA), that corresponds to
the Fig~1 diagram.
Then the experimental cross section was multiplied by
the factor $\sigma^{th}_{Born}/\sigma^{th}_{cut}$.
The results of this rescaling procedure are enlisted in the Tables~3a, 3b.\\

\boldmath
{\large $\sigma_6^{sd}$}. 
\unboldmath
~Low energy data ($2m_\pi < E_{cm} < 2$ GeV).
The treatment of these data is discussed in the following section.

\subsection{Low energy data treatment ($2m_\pi < E_{cm} < 2$ GeV)}
Our treatment of the cross section, 
\boldmath
$\sigma_6^{sd}$
\unboldmath
, data in low energy
range $2m_\pi < \sqrt{s} < 2$ GeV requires special consideration.

All the cross sections $\sigma{}(e^+{}e^-\to hadrons)$
in this range are obtained via  exclusive channel
summation and therefore give only a lower estimate of
the total hadronic cross section.
Below 1 GeV we summed up the $2\pi$ and $3\pi$ channels
from the references
~\cite{Koop:1979??} -- \cite{Akhmetshin:1999uj}
using a  linear
interpolation of the individual data sets within specific energy regions and
combined them  to give the total hadron cross sections in these regions.
The errors were calculated
according to these interpolation and summation procedures.

Such an approach works well if all the data are evenly distributed over $\sqrt{s}$
and have comparable errors.
Otherwise if in the given $\sqrt{s}$ interval
there are few  points of the leading channel with large errors,
and  this interval is filled more densely by the points of less contributing
channels, resonance-like false structures in the total cross section
may arise.

The possibility of such false structures is likely for the exclusive channel data
in the range $1.4 < \sqrt{s} < 2$ GeV,
where we summed the contributions from the channels 
$\ge 3hadrons$, $\pi^+\pi^-$, $K^+{}K^-$, and $K_S K_L$.
To avoid these false structures  we have
determined the sum of  exclusive channels
not for all points, where at least one channel is measured,
but in a fewer number of points, in which the channels yielding 
the major contributions are measured.
In other aspects the summation procedure was the same as the one 
used for the data below 1 GeV.
The exclusive channel data for  $1.4 < \sqrt{s} < 2$ GeV
are ~\cite{Augustin:1983ix} -- \cite{Dolinsky:1991vq}.
The exclusive sum data in the $0.81 < \sqrt{s} < 1.4$ GeV region
are taken from ~\cite{Dolinsky:1991vq}.  
The data used for the exclusive channel summation
around $\phi$ resonance ($0.997 < \sqrt{s} < 1.028$ GeV) 
were corrected by their authors for the initial state radiation,
electronic vertex loops and leptonic 1-loop insertions
insertions into the $\gamma$ propagator.
We have properly rescaled these data points to the Fig~1 diagram.
In the remaining $0.81$--$1.4$ GeV data their
authors applied 
QED-corrections using the Bonneau and Martin~\cite{Bonneau:1971mk} 
prescription and all these data points were rescaled as in the
{\boldmath $\sigma_2^{sd}$\unboldmath}-case.
\\
%
%
%
%
%
%
%
%
%
%
\\
\\
\noindent

\section{Data compilations on {$\sigma^{sd}$} and {$R^{bare}$} } 

After the data selection and rescaling, described in Section~1,
we assemble the complete data set of the total
hadronic cross sections {\boldmath $\sigma^{sd}$\unboldmath}, normalized to the contribution
of the Fig.~1 diagram in accordance with the symbolic relation
{\boldmath $$\sigma^{sd} = \sigma_1^{sd} \cup \sigma_2^{sd} \cup \sigma_5^{sd}
\cup \sigma_6^{sd} \cup \left [ ( R_3^{bare} \cup R_4^{bare}) \cdot
\sigma^{\mu\mu}_{QED~pole,~running~\alpha_{QED}}\right ]$$ \unboldmath}
and the complete data set for the $R$-ratios\\
{\boldmath $$ R^{bare} ={{ \sigma(e^+ e^- \to hadrons)_{Fig.1~diagram}} \over 
{\sigma(e^+ e^- \to \mu^+ \mu^-)_{QED~pole,~running~\alpha_{QED}}}}.$$ \unboldmath}

\noindent
As decribed above, in creating compilations which are uniform in the sense of
a standardized implementation of pure QED radiative corrections to
the ``raw" published experimental data, the values of the running 
$\alpha_{QED}(s)$ and $\Delta\alpha_{QED}^{e}(s)$ were used. 
The procedures for obtaining 
their numerical values and estimates for their uncertainties are described 
in the Appendix. Both compilations are stored and maintained in the PPDS CS 
database. Each record in the compilation corresponds to the record of the 
original data but contains the rescaled data points.
Brief descriptions of the applied conversion procedures 
are stored in the special comment in each record.
The {\boldmath $\sigma$\unboldmath} and {\boldmath $R^{bare}$\unboldmath} 
compilations obtained by this procedure are shown
in the Figs.~5 and 6, respectively.

\section{The continuum regions}
 
To be able to pick out the region where the parton model with QCD corrections can
be tested we need to consider the relevant theoretical formulae.
\subsection{Theoretical relations.}
~In the parton model, before QCD corrections are applied,
 the $R$-ratio is given by
\begin{equation}\label{R0}
R = 3 \sum_q R^0_q = 
3 \sum_q \left [ \beta_q (1 + \frac{1}{2}(1-\beta_q^2))\cdot R_q^{VV} 
+ \beta_q^3 R_q^{AA}\right ]~,
\end{equation}
where
\begin{eqnarray}\label{RVV_RAA}
R_q^{VV} &=&
e_e^2 e_q^2 + 2e_e e_q \overline{v}_e\overline{v}_q~ \mbox{Re}\chi 
+ (\overline{v}_e^2 + \overline{a}_e^2)\overline{v}_q^2~ \vert\chi\vert^2
\nonumber\\
R_q^{AA} &=&
(\overline{v}_e^2 + \overline{a}_e^2)\overline{a}_q^2~ \vert\chi\vert^2
\end{eqnarray}
\noindent
with
\begin{eqnarray}
\chi &=& \frac{1}{16\overline{s}^2\overline{c}^2}
\frac{s}{(s - M_Z^2 + i M_Z\Gamma_Z)}
\nonumber\\
\overline{v} &=& \sqrt{\rho}~ (T_3 - 2Q\overline{s}^2)
\nonumber\\
\overline{a} &=& \sqrt{\rho}~ T_3
\nonumber\\
\rho &=& 1 + \frac{3\sqrt{2}G_F}{16\pi^2} m_t^2 + \cdots~~.
\end{eqnarray}
Here $\overline{s}^2, \overline{c}^2$ are an effective $\sin^2\theta_W, 
\cos^2\theta_W$  defined through renormalized couplings at $s = M_Z^2$:
\begin{equation}
M_Z^2 = \frac{\pi \cdot \alpha_{QED}(M_Z^2)}
{\sqrt{2}G_F \cdot \rho \cdot \overline{s}^2 \cdot \overline{c}^2}~.
\end{equation}
The dominant correction term in $\rho$ originates from $t$-quark loops
in $W$ and $Z$ propagators which result in an $SU(2)$ violation due to
the large mass splitting between the $b$ 
and $t$ quarks ~\cite{Veltman:1977kh}.

Including QCD loops the expression for  $R$ is now (see, e.g.~\cite{Chetyrkin:1996ia},
\cite{Chetyrkin:2000zk}):

\begin{equation}
\begin{array} {rcl}
R &=& 3\sum R_q \\
  &=& 3\sum \left[ R_q^0  +  \left( {\displaystyle \frac{\alpha_s} {\pi}}
      + \left( {\displaystyle
      \frac{\alpha_s} {\pi}} \right)^2 (1.9857-0.1153 N_f) \right. \right. 
\nonumber \\
  &-& \left. \left. \left( {\displaystyle \frac{\alpha_s} {\pi}} \right)^3 
 (6.6368 + 1.2001 N_f + 0.0052 N_f^2 + 1.2395 (\sum Q_q)^2 ) \right) \right. 
\times \nonumber \\
&\times& \left. \left. \left. (f_1 R_q^{VV} + f_2 R_q^{AA}) 
         \right. \right. \right ] 
\end{array}
\label{R}
\end{equation}

\noindent
The coefficients $f_1$ and $f_2$ depend only on the quark velocity 
$\beta = (1- 4M_q^2/s)^{1/2}$, where $M_q$ is the flavour production
threshold mass of the quark $q$.
Schwinger~\cite{Schwinger:1970??} calculated $f_1$ in the QED context and 
parameterised it in a form which is accurate enough for our purposes:

\begin{equation}
f_1 = \beta_q \left ( 1 + \frac{1}{2}(1-\beta_q^2) \right ) 
\frac{4\pi}{3}\left ( \frac{\pi}{2\beta_q} - \frac{3+\beta_q}{4} 
\big{(}\frac{\pi}{2} - \frac{3}{4\pi}\big{)}\right )
\end{equation} 

Here $N_f$ is a number of active quark flavours.
The coefficient $f_2$ (which have no counterpart in QED) were
calculated by Jersak et al~\cite{Jersak:1981uv}.
As there is no compact analytical expression for $f_2$,
we parameterised it as 
\begin{equation}\label{f2}
f_2 = a_4\beta_q^4 + a_3\beta_q^3 + (1 - a_4 - a_3) \beta_q^2
\end{equation}
with $a_4 = -16$, $a_3 = 17$.
>From equations~(\ref{R0}), (\ref{RVV_RAA}), (\ref{R}) one can easily
estimate that even assuming a 100\% error for $f_2$, 
the relative error of $R$ is less than $10^{-7}$ at $\sqrt{s} = 10$ GeV
and less than 0.5\% at the $Z$ pole.
As we did not used $Z$ pole data in our fits,
it could not be a source of large theoretical errors.  

In the massless quark limit, $f_1 = f_2 = 1$.
However, at $\sqrt{s} = 35$ GeV for $b$ quarks one has $\beta = 0.963$
whence $f_1\simeq 1.3$ and $f_2\simeq 1.7$.
Indeed, such a parameterisation of mass effects is valid in QCD only for 
${\cal O}(\alpha_s)$ order. Now the correct parameterisation is known upto
${\cal O}(\alpha_s^3)$ order (\cite{Chetyrkin:1996ia},\cite{Chetyrkin:2000zk}),
 but it was not implemented yet into our programme. 
It is likely that it results in large enough 
discrepancies just in the region where $\alpha_s/\pi$ becomes large. 

The following three-loop parameterisation of $\alpha_s$ 
was chosen here (see, e.g., \cite{Marshall:1989ri}):
\begin{eqnarray}
\ln\big{(}Q^2/\Lambda^2_{\overline{MS}}\big{)} &=& 
\frac{4\pi}{\beta_0\alpha_s}~
- ~\frac{1}{2}\frac{\beta_1}{\beta_0^2}
\ln\left [ \left ( \frac{4\pi}{\beta_0\alpha_s}\right)^2 
          + \frac{\beta_1}{\beta_0^2}
            \left ( \frac{4\pi}{\beta_0\alpha_s}\right )
          + \frac{\beta_2}{\beta_0^3}\right ]
\nonumber\\
&& \phantom{\frac{4\pi}{\beta_0\alpha_s}~}
 - ~\frac{1}{\Delta}\left (\frac{\beta_1}{\beta_0^2}\right )
    \tan^{-1} \left [ \frac{1}{\Delta} \left ( \frac{\beta_1}{\beta_0^2}
   + \frac{2\beta_2}{\beta_0^3}
      \left ( \frac{\beta_0\alpha_s}{4\pi}\right ) \right ) \right ]
\end{eqnarray}
with
\begin{eqnarray}
\beta_0 &=& 11 - 2N_f/3~,
\nonumber\\ 
\beta_1 &=& 2(51 - 19N_f/3)~,
\nonumber\\ 
\beta_2 &=& (2857 - 5033N_f/9 + 325N_f^2/27)/2~, 
\nonumber\\ 
\Delta  &=& \sqrt{4\beta_2/\beta_0^3 - \beta_1^2/\beta_0^4} ~.
\nonumber\\ 
\end{eqnarray}

Here $N_f$ is a number of quark flavours, contributing to the
$\alpha_s$ evolution. The onset of a new flavour $q$ 
takes place at $Q^2 = \mu^2_q = 4 m_q^2$, where, in general, $m_q\neq M_q$.
Here are five 
different $\Lambda_{\overline{MS}}(N_f)$ corresponding to $N_f = 2,3,4,5,6$
above the appropriate quark thresholds.
To sew up $\alpha_s$ at the thresholds an apparent matching condition 
for $\Lambda$'s was used.
Thus at any $Q^2$ $\alpha_s(Q^2)$ is determined by, say, $\alpha_s(M_Z^2)$, 
which has been used as a parameter for fits.

We solved this equation at given $Q^2$ numerically by rewriting it
in the form $x = f(x)$, where $x = 4\pi/(\beta_0\alpha_s)$.
It enabled us to obtain correct $\alpha_s$ even at $Q^2/\Lambda^2\sim 1$, 
where the expansion by powers of leading logarithms 
(see, e.g.,\cite{Hinchliffe:2000yq})
is no more valid. At $Q^2/\Lambda^2\gg 1$ both methods yield the same result.
(See Fig.~8)

\subsection{Preliminary fit results}
In order to check the consistency of the description
of the data on
$$R = \sigma(e^+{}e^-\to hadrons)/\sigma(e^+{}e^-\to\mu^+\mu^-)$$
with the theoretical $R$-ratio expression
we have performed several fits using a sub-set of the total compiled
data set of $R$ which is applicable to the perturbative domain.
This sub-set of the data is defined by the the following steps:
\begin {enumerate}
\item  The low energy region $2m_\pi < \sqrt{s} < 2$ GeV is completely excluded;
\item All the data above 70 GeV are excluded, as we did not attempt
to perform any fits of $Z$ pole parameters $M_Z$, $\Gamma_Z$,
$\overline{s}^2_W$ and $M_{top}$. This exclusion justified because these 
data slightly depend on $\alpha_s(M_Z)$ and quark masses and thus their 
influence to the fits of these parameters would be rather negligable.
\item Data 
located closer than 20 Breit--Wigner widths
to the narrow hadronic $1^{--}$ resonances $J/\psi(1S)$, $\psi(2S)$, $\psi(3770)$,
$\Upsilon(1S)$, $\Upsilon(2S)$, $\Upsilon(3S)$, $\Upsilon(4S)$,
$\Upsilon(10860)$ and $\Upsilon(11020)$ are completely excluded;
\item We also completely exclude SLAC-SPEAR-MARK-1 
data~\cite{Siegrist:1982zp} in the $2.6$ GeV $< \sqrt{s} < 7.8$ GeV range.
These data systematically lie $\simeq$15\% above 
other experiments in the same interval.
\end{enumerate}

Some remarks on our  treatment of the quark masses  
should be made here.
We distinguished the $\gamma^*\to q\overline{q}$ production 
threshold masses $M_q$
and the QCD quark masses $m_q$. 
The $M_q$'s should be set actually equal to the masses 
of the lightest mesons, above which pair production threshold
the onset of the new flavour $q$ gives rise to the 
continuum cross section $\sigma(e^+{}e^-\to hadrons)$.
The latters are  just
$m_q = \mu_q/2$, where $\mu_q$ is the energy scale 
at which the onset of the new flavour $q$ in the 
evolution of $\alpha_s$ takes place. 
We fitted only $M_c$ and $M_b$, the other $M_q$ and $m_q$ were 
fixed at their central PDG values.

The standard $\chi^2$ of the least squares method with weights 
as inverse squared total experimental errors (neglecting 
correlations in data) is used with the 
standard MINUIT\cite{James:1975dr} package.

After performing the exclusions 1) -- 4) 
we fixed $\alpha_s(M_Z)$, $M_Z$, $\Gamma_Z$, $\overline{s}^2_W$,
$M_q$ ($q = u,d,s,t$), $m_q$ ($q = u,d,s,c,b,t$)
at the values shown in the Table~1, leaving as 
free parameters just $M_c$ and $M_b$.
Another essential feature of our fit ({\bf I})
was that we retained as a free parameters
left and right boundaries of the interval $[\sqrt{s_1},\sqrt{s_2}]$,
$3.0\le\sqrt{s_1}\le 3.670$ GeV, $3.870\le\sqrt{s_2}\le 5.0$ GeV,
which was to be excluded from the 
data set in the process of $\chi^2/dof$ minimization itself.\footnote{Narrow
$\psi$ family resonances between 3 and 4 GeV were excluded by default.
Somewhat controversial is the question, whether the gaps between $\psi$'s
can be really treated as the continuum regions.
$R$ ratio demonstrates no step up until $\sqrt{s} = 3.9-4.0$ GeV, just at the 
left knee of the broad $\psi(4040)$ resonance, where the continuum
approximation of $R$ does not work. 
Nevertheless, this step results in the fitted $c$-quark mass
$M_c = 1.971$ (see fit ({\bf III}) in the Table~1).

In pQCD fits, performed in Refs.\ \cite{Martin:1995we},~\cite{Swartz:1996hc} 
the interval $3.0 < \sqrt{s} < 4.0$ GeV with the resonances excluded
was treated as a continuum.}
Thus, the number of degrees of freedom was also variable
during this fit.
Such a $\chi^2/dof$ minimization procedure
cancelled possible arbitrarities in the exclusions
of broad resonances in the region above $c\overline{c}$ threshold.

In the fit ({\bf II}) we released $\alpha_s(M_Z)$, fixing at the same time 
the excluded region $\sqrt{s} = 3.09 \div 4.44$ GeV.
$M_c$, $M_b$ were fixed at their fitted values obtained in the fit ({\bf I}). 
Fit ({\bf II}) results in too high value $\alpha_s(M_Z) = 0.128\pm 0.032$.


Preliminary fit parameter settings and 
results are enlisted in the Table~1.
The reduced data set is shown on the Fig.~7.
The references used for our fit and 
the list of excluded $\sqrt{s}$ regions containing hadronic 
resonances are quoted in the Table~2.


\section{Summary. Assessed Data Compilations.}

In summary:
\begin{itemize}
\item 
We have created two complementary computerized ``raw" data 
compilations on $\sigma$ and $R$ with data presented as in the original 
publications in all cases except the low energy subsample. In the
low energy region,
where there are no direct measurements of the total cross 
section, we obtain estimates of the total cross section either as the 
sum of exclusive channels or as the sum of the two-body exclusive 
channels with the data on $e^+~e^- \to \ge 3 hadrons$.  
\item On the basis of above two data sets we have created
compilations of data on {\boldmath $\sigma^{sd}$\unboldmath} (corrected to the level of Fig~1) and
{\boldmath $R^{bare}$\unboldmath} (corrected to the level of Fig.~2) with one-to-one
correspondence between the data points. Where necessary the data have been rescaled to the
standard style of implementing the pure QED radiative corrections to the initial
state and to photonic propagator to produce a data set which is suitable for tests of the parton
model with pQCD corrections for {\boldmath $R^{bare}$\unboldmath}, and to be able to obtain more reliable
estimates for $\Delta\alpha_{QED}^{had}(Q^2)$.
\item  From the total data compilation on {\boldmath $R^{bare}$\unboldmath} we have defined a
``continuum" data compilation sub-set which can be used in conjunction with other
data in the refinements of the $\alpha_s(Q^2)$ evolution and possibly in 
the global fits of the Standard Model.
\end{itemize}

All data files are accessible by: \\
\begin{center}
{\rm http://wwwppds.ihep.su:8001/comphp.html}
\end{center}

and will be accessible from the PDG site soon.

\section*{Acknowledgements}

We are pleased to thank: R.Yu.~Pirogov (for the participation 
in the early stage of the work); Yu.F.~Pirogov for the discussions of the 
theory aspects;
S.R.~Slabospitsky, the COMPAS ReacData manager for his kind help; D.E.~Groom 
and R.M.~Barnett for 
the fruitful discussions and encouragements; Z.G.~Zhao and Y.S.~Zhu for
providing us the BES numerical data,
and finally to V.S.~Lugovsky for 
providing us the dynamical Web graphic access to CS database. 
The HEPDATA database project is funded by a grant from the PPARC(UK).
 
\section{Appendix. Running $\alpha_{QED}$ and $\Delta \alpha^{had}_{QED}$ }
As mentioned above, the authors of the original papers
have published data in different forms which required the calculation of
several types of factors, containing $\alpha_{QED}(s)$,  
to rescale the data to, say, ``bare'' $R$-ratios.
$\alpha_{QED}(s)$ can be expressed in the form 
(see: \cite{Cabibbo:1960??},
\cite{Cabibbo:1961sz}, \cite{Nevzorov:1994tt}, \cite{Eidelman:1995ny})

\begin{eqnarray}\label{alpha_QED}
\alpha_{QED}(s) &=& \frac{\alpha(0)}{1 - \Delta \alpha (s)}~,
\nonumber\\
\Delta \alpha (s) &=& \Delta \alpha^{had} (s) + \Delta \alpha^{lep} (s)~,
\end{eqnarray}
where $\Delta \alpha^{had}(s)$, $\Delta \alpha^{lep} (s)$
are hadronic and leptonic contributions to the QED vacuum polarization,
respectively.

$\Delta \alpha^{lep} (s)$ is well determined in perturbative QED
as a sum of leptonic loop contributions,
%
\begin{equation}\label{Delta_alpha_lep}
\Delta\alpha^{lep} (s) = \sum_{l=e,\mu,\tau} \frac{\alpha(0)}{3\pi}
\left[ \ln\frac{s}{m_l^2} - \frac{5}{3} + {\cal O}\left(\frac{m_l^2}{s}\right)\right]~.
\end{equation}

The situation with $\Delta \alpha^{had} (s)$ is more complicated 
due to an essentially non-perturbative character of the strong interaction
at low energy scales.  Using the unitarity condition and the analyticity of the
scattering amplitudes, one can express $\Delta \alpha^{had} (s)$
in the form of a subtracted dispersion relation
(see, e.g., \cite{Nevzorov:1994tt}, \cite{Eidelman:1995ny})
\begin{equation}\label{disp_rel}
\Delta\alpha^{had} (s) = - \frac{\alpha(0) s}{3\pi}
\int_{4m_\pi^2}^{\infty} \frac{R(s') ds'} {s' (s' - s - i0)}~,
\end{equation}
where $R(s)$ is the ``bare'' hadronic $R$-ratio.

This relation allows to effectively combine for its evaluation
the $pQCD$ $R$-ratios in the continuum $\sqrt{s}$ intervals and
experimental $R$ data in the non-perturbative ones.

We have evaluated the dispersion integral as
\begin{equation}\label{disp_rel_comb}
\Delta\alpha^{had} (s) = - \frac{\alpha(0) s}{3\pi}
\left [ 
\int_{4m_\pi^2}^{\sqrt{s_{cut}}} \frac{R^{data}_{bare}(s') ds'} {s' 
(s' - s - i0)} +
\int_{\sqrt{s_{cut}}}^{\infty} \frac{3\sum_q Q^2_q ds'} {s' 
(s' - s - i0)} \right ]~,
\end{equation}
where the first integral was calculated numerically as a trapezoidal sum over 
the weighted average of experimental $R$ points in the range 
$2m_\pi < \sqrt{s'} < \sqrt{s_{cut}} = 19.5$ GeV\footnote{Choosing 
$S_{cut}$ we must satisfy at least the following 
two requirements:\\
1)~ $S_{cut}$ should lie well above all the hadronic resonances, i.e.
1-loop perturbative QED can be used to calculate hadronic vaquum
polarization at $s > s_{cut}$;\\
2)~ there should be enough densely distributed experimental points at 
$s < s_{cut}$ for the trapezoidal 
numerical evaluation of the dispersion 
integral over $2m_\pi < \sqrt{s} < \sqrt{s_{cut}}$.

As there are few experimental points in the interval
$13 < \sqrt{s} < 30$ GeV, $\sqrt{s_{cut}} = 19.5$ GeV
appears to be a compromise between these requirements.}

Our numerical evaluation procedure is in general similar to the one 
applied in Ref.~\cite{Eidelman:1995ny}, except that we performed the 
trapezoidal integration over all 
resonances, except
$\phi(1020)$.\footnote{As the calculation of
$R^{data}_{bare}$ from the original $\sigma$ and $R$ data
(see subsection 1.1) in turn requires the evaluation
of $\Delta\alpha^{had}(s)$, we applied the following simple
iterative procedure. 

Taking as a zero approximation $R(s)$, obtained from the original data
using  1-loop perturbative $\alpha_{QED}(s)$,
we evaluate $\Delta \alpha^{had(1)}(s)$.
Using the latter, we compute the  next approximation
of $R(s)$ from the original data,
the evaluate the integral again to obtain $\Delta \alpha^{had(2)}(s)$, 
and so on.
This process converges well even in resonance regions after 3 iterations.
We restricted ourselves to 5 iterations. (For such a number
the finite machine precision is likely to result in the error, much less
than the error of the method itself).}

We obtain here $\Delta \alpha^{had}(M_Z^2) = 0.0271\pm 0.0004$(exp.),
being consistent with the latest known to us results
$0.027382 \pm 0.000197$ and $0.027612 \pm 0.000220$, obtained with two
methodically different low energy data sets in Ref.\ \cite{Martin:2000by}.

The behaviour of $\Delta\alpha^{lep}$, 
$\Delta\alpha^{had}_{(1-loop~~QED)}$ and 
$\Delta\alpha^{had}_{(dispersion)}$ 
is depicted in Fig.~9.

\begin{table}[htbp]
\paragraph{Table 1.} Fit results in the defined ``continuum" regions. 
Adjustable parameters are shown in bold.
\vspace{0.5cm}

\begin{center}
\begin{tabular}{|c|c|c|c|}
\hline
&&&\\[-2mm]
Parameter&{\bf I}& {\bf II} & {\bf III} \\[2mm]
\hline
$\alpha_s(M_Z^2)$ & 0.1181 
                  & {\bf 0.128(32) }  
                  &{\bf 0.126(37)}
\\
\hline
$M_Z$ & 91.187 & 91.187 & 91.187\\
\hline
$\Gamma_Z$ & 2.4944 & 2.4944 & 2.4944\\
\hline
$\overline{s}^2$ & 0.23117 & 0.23117 & 0.23117\\
\hline
$M_t$ & 174.3 & 174.3 & 174.3\\
\hline
$M_u$ & 0.140 & 0.140 & 0.140\\
$M_d$ & 0.140 & 0.140 & 0.140\\
$M_s$ & 0.492 & 0.492 & 0.492\\
\hline
{\bf $M_c$} & {\bf 1.500(18) } 
            & 1.5 
            & {\bf 1.9710(1)} 
\\ 
{\bf $M_b$} & {\bf 5.23   $\pm$ 1.16 } 
            & 5.23
            & {\bf 6.0 $\pm$ 1.4}
\\
\hline
$m_u$ & 0.003 & 0.003&0.003\\
$m_d$ & 0.006 & 0.006 &0.006\\
$m_s$ & 0.120 & 0.120 &0.120\\
$m_c$ & 1.2   & 1.2  & 1.2\\
$m_b$ & 4.2  & 4.2 &4.2\\
$m_t$ & 174.3 & 174.3&174.3\\
\hline \hline
		    &2$m_\pi \div$ 2.0
		    &2$m_\pi \div$ 2.0
		    &2$m_\pi \div$ 2.0
\\	
                    &   3.093$\div$3.113 
		    &	
		    & 3.093$\div$3.113
\\
		    &
                    &
		    & 3.684$\div$3.688
\\
            	    &
		    &
                    & 3.670$\div$3.870
\\ \cline{2-4}
{\bf Excluded}      &   3.175(247) $\div$\phantom{\bf 4.44}
                    &   3.09 $\div$4.44
                    &   4.000$\div$4.400
\\
                    &   {\bf 4.319(105)}   
                    &   
                    &   
\\ \cline{2-4}
   $\sqrt{S}$       &   9.450$\div$9.470 
                    &   9.450$\div$9.470 
                    &  9.450$\div$9.470
\\
                    &   10.000$\div$10.025 
                    &   10.000$\div$10.025
                    & 10.000$\div$10.025
\\
{\bf intervals}     &   10.34 $\div$10.37  
                    &   10.34 $\div$10.37 
                    &  10.34 $\div$10.37
\\
                    &   10.52 $\div$10.64 
                    &  10.52 $\div$10.64
                    &  10.52 $\div$10.64
\\
                    &   10.75 $\div$10.97 
                    &  10.75 $\div$10.97
                    &  10.75 $\div$10.97
\\
                    &   11.00 $\div$11.20 
                    &  11.00 $\div$11.20
                    &  11.00 $\div$11.20
\\
                    &  70 $\div$ 188.7
                    &  70 $\div$ 188.7
                    &  70 $\div$ 188.7
\\
\hline  \hline
&&&\\[-2mm]
{\bf $\chi^2/$dof} & {\bf 0.690} 
                   & {\bf 0.665} 
                   & {\bf 0.822}
\\[2mm] 
\hline
\end{tabular}
\end{center}
\end{table}

\def\c{\cite}

\footnotesize
\begin{table}[htbp]
\paragraph{Table 2.} References of  preliminary candidate 
data to form the continuum domain.

\begin{center} 
\begin{tabular}{|c|c|l|l|l|}
\hline
&&&&\\[-2mm]
Ref. No.     & Data type  & $N_{points}$ & $E_{min}$ & $E_{max}$ 
\\[2mm]
\hline
&&&&\\[-2mm]
\cite{Bai:2001ct}            & R          & 85   & 2.00     & 4.80     \\
\cite{Bartoli:1975jx}        & $\sigma$   & 1    & 2.23     & 2.23     \\
\cite{Augustin:1975yq}       & $\sigma$   & 11   & 2.40     & 5.00     \\
\cite{Bai:2000pk}            & R          & 6    & 2.60     & 5.00     \\
\cite{Rapidis:1977cv}        & R          & 2    & 3.598    & 3.886    \\
\cite{Brandelik:1978ei}      & R          & 33   & 3.6025   & 5.1950   \\
\cite{Osterheld:1986hw}      & R          & 27   & 3.878    & 4.496    \\
\cite{Edwards:1990pc}        & R          & 15   & 5.00     & 7.40     \\
\cite{Blinov:1996fw}         & R          & 31   & 7.30     & 10.29    \\
\cite{Niczyporuk:1982ya}     & R          & 3    & 7.440    & 9.415    \\
\cite{Criegee:1982qx}        & $\sigma$   & 13   & 9.30     & 9.48     \\ 
\cite{Albrecht:1992vp}       & R          & 1    & 9.36     & 9.36     \\
\cite{Jakubowski:1988cd}     & R          & 1    & 9.39     & 9.39     \\
\cite{Albrecht:1982bq}       & R          & 1    & 9.51     & 9.51     \\
\cite{Bock:1980ag}           & R          & 1    & 10.04    & 10.04    \\
\cite{Rice:1982br}           & R          & 1    & 10.43    & 10.43    \\
\cite{Giles:1984vq}          & R          & 1    & 10.49    & 10.49    \\
\cite{Althoff:1984ew}        & R          & 12   & 12.00    & 41.40    \\
\cite{Brandelik:1980bv}      & R          & 7    & 12.00    & 31.25    \\
\cite{Brandelik:1982zi}      & R          & 14   & 12.00    & 36.00    \\
\cite{Barber:1980uq}         & R          & 12   & 12.00    & 35.80    \\
\cite{Naroska:1987si}        & R          & 20   & 12.00    & 46.47    \\
\cite{Adeva:1986nt}          & R          & 18   & 12.00    & 46.47    \\
\cite{Behrend:1987md}        & R          & 9    & 14.00    & 46.60    \\
\cite{Braunschweig:1990yd}   & R          & 4    & 14.03    & 43.70    \\
\cite{VonZanthier:1991we}    & R          & 1    & 29.00    & 29.00    \\
\cite{Fernandez:1985yw}      & R          & 1    & 29.00    & 29.00    \\
\cite{Barber:1979ru}         & R          & 1    & 31.57    & 31.57    \\
\cite{Barber:1982tc}         & R          & 1    & 34.85    & 34.85    \\
\cite{Behrend:1984ub}        & R          & 2    & 34.86    & 42.72    \\
\cite{Althoff:1984us}        & R          & 2    & 41.45    & 44.20    \\
\cite{Adachi:1990yz}         & R          & 12   & 50.00    & 61.40    \\
\cite{Yoshida:1987sf}        & R          & 2    & 50.00    & 52.00    \\
\cite{Kumita:1990my}         & R          & 13   & 50.00    & 61.40    \\
\cite{Abe:1993fy}            & $\sigma$   & 9    & 57.37    & 59.84    \\ 
\cite{Miyabayashi:1995ej}    & $\sigma$   & 1    & 57.77    & 57.77    \\
\cite{Abe:1990rr}            & R          & 2    & 63.60    & 64.00    \\[2mm]
\hline
\end{tabular}
\end{center}
\end{table}
\newpage
\footnotesize
\paragraph{Table 3a.} ZFITTER rescaling results for the $Z$ pole.
\begin{center} 
\begin{tabular}{|c||c|c||c|c|}
\hline
&&&&\\[-2mm]
$\sqrt{s}$, GeV & $\sigma^{th}_{cut}$ [nb]& $\sigma^{th}_{born} [nb]$ &
$\sigma^{exp}$ [mb]& 
$\sigma^{exp} \cdot (\sigma^{th}_{born} / \sigma^{th}_{cut})$ [mb] 
\\[2mm]
\hline
&&&&\\[-2mm]
  88.22300  &   0.44967E+01 &  0.60846E+01 &    0.44800E-05 &  0.60619E-05 \\
  88.22300  &   0.44967E+01 &  0.60846E+01 &    0.46100E-05 &  0.62378E-05 \\
  88.22400  &   0.44992E+01 &  0.60881E+01 &    0.46300E-05 &  0.62651E-05 \\
  88.23101  &   0.45168E+01 &  0.61131E+01 &    0.44600E-05 &  0.60363E-05 \\
  88.27800  &   0.46369E+01 &  0.62843E+01 &    0.50400E-05 &  0.68306E-05 \\
  88.46400  &   0.51600E+01 &  0.70320E+01 &    0.51500E-05 &  0.70183E-05 \\
  88.46400  &   0.51600E+01 &  0.70320E+01 &    0.54700E-05 &  0.74544E-05 \\
  88.48001  &   0.52091E+01 &  0.71020E+01 &    0.52200E-05 &  0.71169E-05 \\
  88.48101  &   0.52121E+01 &  0.71064E+01 &    0.53500E-05 &  0.72944E-05 \\
  89.21701  &   0.83923E+01 &  0.11703E+02 &    0.84100E-05 &  0.11728E-04 \\
  89.22200  &   0.84219E+01 &  0.11746E+02 &    0.84800E-05 &  0.11827E-04 \\
  89.22600  &   0.84458E+01 &  0.11781E+02 &    0.84300E-05 &  0.11759E-04 \\
  89.23600  &   0.85058E+01 &  0.11868E+02 &    0.85200E-05 &  0.11888E-04 \\
  89.28300  &   0.87956E+01 &  0.12290E+02 &    0.96800E-05 &  0.13526E-04 \\
  89.45500  &   0.99723E+01 &  0.14007E+02 &    0.99900E-05 &  0.14032E-04 \\
  89.45500  &   0.99723E+01 &  0.14007E+02 &    0.10010E-04 &  0.14060E-04 \\
  89.47000  &   0.10084E+02 &  0.14170E+02 &    0.10150E-04 &  0.14262E-04 \\
  89.47200  &   0.10099E+02 &  0.14192E+02 &    0.10130E-04 &  0.14235E-04 \\
  90.20800  &   0.17921E+02 &  0.25591E+02 &    0.17860E-04 &  0.25505E-04 \\
  90.21200  &   0.17977E+02 &  0.25672E+02 &    0.18230E-04 &  0.26034E-04 \\
  90.21701  &   0.18047E+02 &  0.25773E+02 &    0.18000E-04 &  0.25706E-04 \\
  90.21701  &   0.18047E+02 &  0.25773E+02 &    0.18590E-04 &  0.26549E-04 \\
  90.22600  &   0.18173E+02 &  0.25956E+02 &    0.18740E-04 &  0.26765E-04 \\
  90.22700  &   0.18188E+02 &  0.25976E+02 &    0.18320E-04 &  0.26165E-04 \\
  90.22800  &   0.18202E+02 &  0.25996E+02 &    0.18210E-04 &  0.26008E-04 \\
  90.23800  &   0.18343E+02 &  0.26201E+02 &    0.18680E-04 &  0.26681E-04 \\
  90.24000  &   0.18372E+02 &  0.26242E+02 &    0.18830E-04 &  0.26896E-04 \\
  90.28400  &   0.19005E+02 &  0.27152E+02 &    0.19560E-04 &  0.27945E-04 \\
  91.03400  &   0.29394E+02 &  0.40836E+02 &    0.29940E-04 &  0.41594E-04 \\
  91.20700  &   0.30374E+02 &  0.41415E+02 &    0.30590E-04 &  0.41710E-04 \\
  91.20800  &   0.30377E+02 &  0.41414E+02 &    0.30100E-04 &  0.41037E-04 \\
  91.21500  &   0.30396E+02 &  0.41404E+02 &    0.30460E-04 &  0.41491E-04 \\
  91.21701  &   0.30401E+02 &  0.41400E+02 &    0.30540E-04 &  0.41590E-04 \\
  91.22200  &   0.30413E+02 &  0.41391E+02 &    0.30330E-04 &  0.41278E-04 \\
  91.22300  &   0.30415E+02 &  0.41389E+02 &    0.30190E-04 &  0.41082E-04 \\
  91.22300  &   0.30415E+02 &  0.41389E+02 &    0.30480E-04 &  0.41477E-04 \\
  91.23001  &   0.30431E+02 &  0.41373E+02 &    0.30440E-04 &  0.41386E-04 \\
  91.23800  &   0.30447E+02 &  0.41352E+02 &    0.30630E-04 &  0.41601E-04 \\
  91.23900  &   0.30448E+02 &  0.41349E+02 &    0.29960E-04 &  0.40686E-04 \\
\hline
\end{tabular}
\end{center}

\newpage
\vspace{-2cm}
\footnotesize
\paragraph{Table 3a (continued).} ZFITTER rescaling results for the $Z$ pole.
\begin{center} 
\begin{tabular}{|c||c|c||c|c|}
\hline
&&&&\\[-2mm]
$\sqrt{s}$, GeV & $\sigma^{th}_{cut}$ [nb]& $\sigma^{th}_{born} [nb]$ &
$\sigma^{exp}$ [mb]& 
$\sigma^{exp} \cdot (\sigma^{th}_{born} / \sigma^{th}_{cut})$ [mb] 
\\[2mm]
\hline
&&&&\\[-2mm]
  91.25400  &   0.30472E+02 &  0.41300E+02 &    0.30450E-04 &  0.41270E-04 \\
  91.25400  &   0.30472E+02 &  0.41300E+02 &    0.30460E-04 &  0.41284E-04 \\
  91.28000  &   0.30494E+02 &  0.41187E+02 &    0.30440E-04 &  0.41113E-04 \\
  91.28900  &   0.30497E+02 &  0.41139E+02 &    0.30860E-04 &  0.41629E-04 \\
  91.29400  &   0.30497E+02 &  0.41111E+02 &    0.30450E-04 &  0.41048E-04 \\
  91.52900  &   0.29641E+02 &  0.38504E+02 &    0.29210E-04 &  0.37945E-04 \\
  91.95201  &   0.25217E+02 &  0.30140E+02 &    0.25310E-04 &  0.30251E-04 \\
  91.95301  &   0.25205E+02 &  0.30119E+02 &    0.24780E-04 &  0.29611E-04 \\
  91.96701  &   0.25030E+02 &  0.29821E+02 &    0.24640E-04 &  0.29356E-04 \\
  91.96900  &   0.25005E+02 &  0.29779E+02 &    0.24690E-04 &  0.29403E-04 \\
  92.20700  &   0.22038E+02 &  0.24922E+02 &    0.21830E-04 &  0.24686E-04 \\
  92.20900  &   0.22014E+02 &  0.24883E+02 &    0.21570E-04 &  0.24382E-04 \\
  92.21500  &   0.21941E+02 &  0.24768E+02 &    0.21220E-04 &  0.23954E-04 \\
  92.22600  &   0.21807E+02 &  0.24558E+02 &    0.22010E-04 &  0.24786E-04 \\
  92.28200  &   0.21135E+02 &  0.23509E+02 &    0.21240E-04 &  0.23626E-04 \\
  92.56200  &   0.18034E+02 &  0.18865E+02 &    0.16660E-04 &  0.17428E-04 \\
  92.95201  &   0.14545E+02 &  0.14006E+02 &    0.14590E-04 &  0.14049E-04 \\
  92.95301  &   0.14537E+02 &  0.13995E+02 &    0.14120E-04 &  0.13594E-04 \\
  92.96601  &   0.14437E+02 &  0.13862E+02 &    0.14440E-04 &  0.13865E-04 \\
  92.96800  &   0.14421E+02 &  0.13841E+02 &    0.14110E-04 &  0.13542E-04 \\
  93.20800  &   0.12739E+02 &  0.11648E+02 &    0.12580E-04 &  0.11503E-04 \\
  93.20900  &   0.12733E+02 &  0.11640E+02 &    0.12480E-04 &  0.11409E-04 \\
  93.22000  &   0.12663E+02 &  0.11551E+02 &    0.12330E-04 &  0.11248E-04 \\
  93.22800  &   0.12612E+02 &  0.11487E+02 &    0.12380E-04 &  0.11276E-04 \\
  93.28601  &   0.12254E+02 &  0.11035E+02 &    0.11770E-04 &  0.10599E-04 \\
  93.70100  &   0.10097E+02 &  0.84120E+01 &    0.10200E-04 &  0.84981E-05 \\
  93.70201  &   0.10092E+02 &  0.84067E+01 &    0.10070E-04 &  0.83883E-05 \\
  93.71601  &   0.10030E+02 &  0.83341E+01 &    0.10100E-04 &  0.83921E-05 \\
  93.71701  &   0.10026E+02 &  0.83289E+01 &    0.99500E-05 &  0.82660E-05 \\
  94.20201  &   0.82093E+01 &  0.62771E+01 &    0.78200E-05 &  0.59795E-05 \\
  94.20201  &   0.82093E+01 &  0.62771E+01 &    0.79900E-05 &  0.61094E-05 \\
  94.21900  &   0.81557E+01 &  0.62189E+01 &    0.78800E-05 &  0.60087E-05 \\
  94.22300  &   0.81432E+01 &  0.62053E+01 &    0.80600E-05 &  0.61420E-05 \\
  94.27700  &   0.79772E+01 &  0.60261E+01 &    0.75900E-05 &  0.57336E-05 \\
  95.03601  &   0.61405E+01 &  0.41416E+01 &    0.64400E-05 &  0.43436E-05 \\
\hline
\end{tabular}
\end{center}

\newpage
\small
\paragraph{Table 3b.} ZFITTER rescaling results for the LEP II--III data.

\begin{center} 
\begin{tabular}{|c||c|c||c|c|}
\hline
&&&&\\[-2mm]
$\sqrt{s}$, GeV & $\sigma^{th}_{cut} [nb]$ & $\sigma^{th}_{born} [nb]$ &
$\sigma^{exp}$ [mb]& 
$\sigma^{exp} \cdot (\sigma^{th}_{born} / \sigma^{th}_{cut})$ [mb]
\\[2mm]
\hline
&&&&\\[-2mm]
 130.00000  &   0.82691E-01 &  0.82628E-01 &    0.84200E-07 &  0.84136E-07 \\
 130.11999  &   0.82294E-01 &  0.82264E-01 &    0.79700E-07 &  0.79670E-07 \\
 130.19999  &   0.82031E-01 &  0.82022E-01 &    0.82100E-07 &  0.82091E-07 \\
 130.19999  &   0.82031E-01 &  0.82022E-01 &    0.71600E-07 &  0.71592E-07 \\
 136.08000  &   0.66194E-01 &  0.67219E-01 &    0.66700E-07 &  0.67732E-07 \\
 136.10000  &   0.66150E-01 &  0.67177E-01 &    0.66600E-07 &  0.67634E-07 \\
 136.19999  &   0.65930E-01 &  0.66967E-01 &    0.65100E-07 &  0.66124E-07 \\
 136.19999  &   0.65930E-01 &  0.66967E-01 &    0.58800E-07 &  0.59725E-07 \\
 161.30000  &   0.34218E-01 &  0.35777E-01 &    0.37300E-07 &  0.38999E-07 \\
 161.30000  &   0.34218E-01 &  0.35777E-01 &    0.40900E-07 &  0.42763E-07 \\
 161.30000  &   0.34218E-01 &  0.35777E-01 &    0.29940E-07 &  0.31304E-07 \\
 172.10000  &   0.28336E-01 &  0.29787E-01 &    0.30300E-07 &  0.31852E-07 \\
 172.10000  &   0.28336E-01 &  0.29787E-01 &    0.26400E-07 &  0.27752E-07 \\
 172.30000  &   0.28244E-01 &  0.29693E-01 &    0.28200E-07 &  0.29646E-07 \\
 182.69000  &   0.24017E-01 &  0.25342E-01 &    0.23700E-07 &  0.25007E-07 \\
 182.69999  &   0.24014E-01 &  0.25338E-01 &    0.21710E-07 &  0.22907E-07 \\
 182.69999  &   0.24014E-01 &  0.25338E-01 &    0.24700E-07 &  0.26062E-07 \\
 188.63000  &   0.22095E-01 &  0.23352E-01 &    0.22100E-07 &  0.23358E-07 \\
 188.69999  &   0.22073E-01 &  0.23330E-01 &    0.23100E-07 &  0.24415E-07 \\
\hline
\end{tabular}
\end{center}

\newpage
\normalsize

\def\PL{Phys.\ Lett.\ }
\def\PRL{Phys.\ Rev.\ Lett.\ }
\def\PRPL{Phys.\ Rep.\ }
\def\ZP{Zeit.\ Phys.\ }
\def\YF{Yad.\ Fiz.\ }
\def\PR{Phys.\ Rev.\ }
\def\NP{Nucl.\ Phys.\ }
\def\NC{Nuovo\ Cim.\ }
\def\LNC{Nuovo\ Cim.\ Lett.\ }
\def\EPJ{Europ.\ J.\ Phys.\ }
\def\SJNP{Sov.\ J. of Nucl.\ Phys.\ }
\def\NOVO{Budker INP Preprint }
\def\BUDKERINP{Budker INP Preprint }
                      
\def\et{{\it et al.}}

\vspace{1ex}
\begin{flushright}
{\it Received 16 May, 2001}
\end{flushright}
\newpage

Figure Captions

\paragraph{Fig.~3:}{
{World data on the 
total cross section of
$e^+{}e^-\to hadrons$ as presented in the original publications.}
}\\

\paragraph{Fig.~4:~}{ { 
World data on the experiment-to-theory ratio $R = \sigma{}(e^+{}e^-\to 
{hadrons})_{experimental}/\sigma{}(e^+{}e^-\to\mu^+\mu^-)_{QED~simple~pole}$ 
 as presented in the original publications.}}\\

\paragraph{Fig.~5:~}{
World data on the 
total cross section of
$e^+{}e^-\to hadrons$.
Curves are an educative guide.
Solid curve is the cross section prediction 
obtained in the three-loop
QCD approximation with the effect of non-zero quark masses
taken into account.}\\

\paragraph{Fig.~6:~}{$R$-ratio. 
Data set is the same as on the Fig.~5. 
Solid curve is the $R$-ratio prediction obtained in the three-loop
QCD approximation with the effect of non-zero quark masses
taken into account.
Dashed curve is a ``naive" quark parton model prediction
for the $R$-ratio}\\

\paragraph{Fig.~7:~}{Data set 
used for the fits.}

\paragraph{Fig.~8:~}{A comparison of $R$-ratio
obtained with $\alpha_s(s)$, evaluated by our numerical method (solid curve)
and by the method described in \cite{Hinchliffe:2000yq} (dashed curve).
}

\paragraph{Fig.~9:~}{The behaviour of $\Delta\alpha_{QED}$.
 $\Delta\alpha_{had}^{(QED)}$  is the quark contribution to 
$\Delta\alpha_{(QED)}$ calculated in 1-loop QED.
$\Delta\alpha_{had}^{(disp)}$ is a 
hadronic contribution to $\Delta\alpha_{(QED)}$ calculated
using the dispersion integral over $\sigma_{had}(s)$, 
$2m_\pi < \sqrt{s} < 20$ GeV, and 1-loop QED above 20 GeV.
}

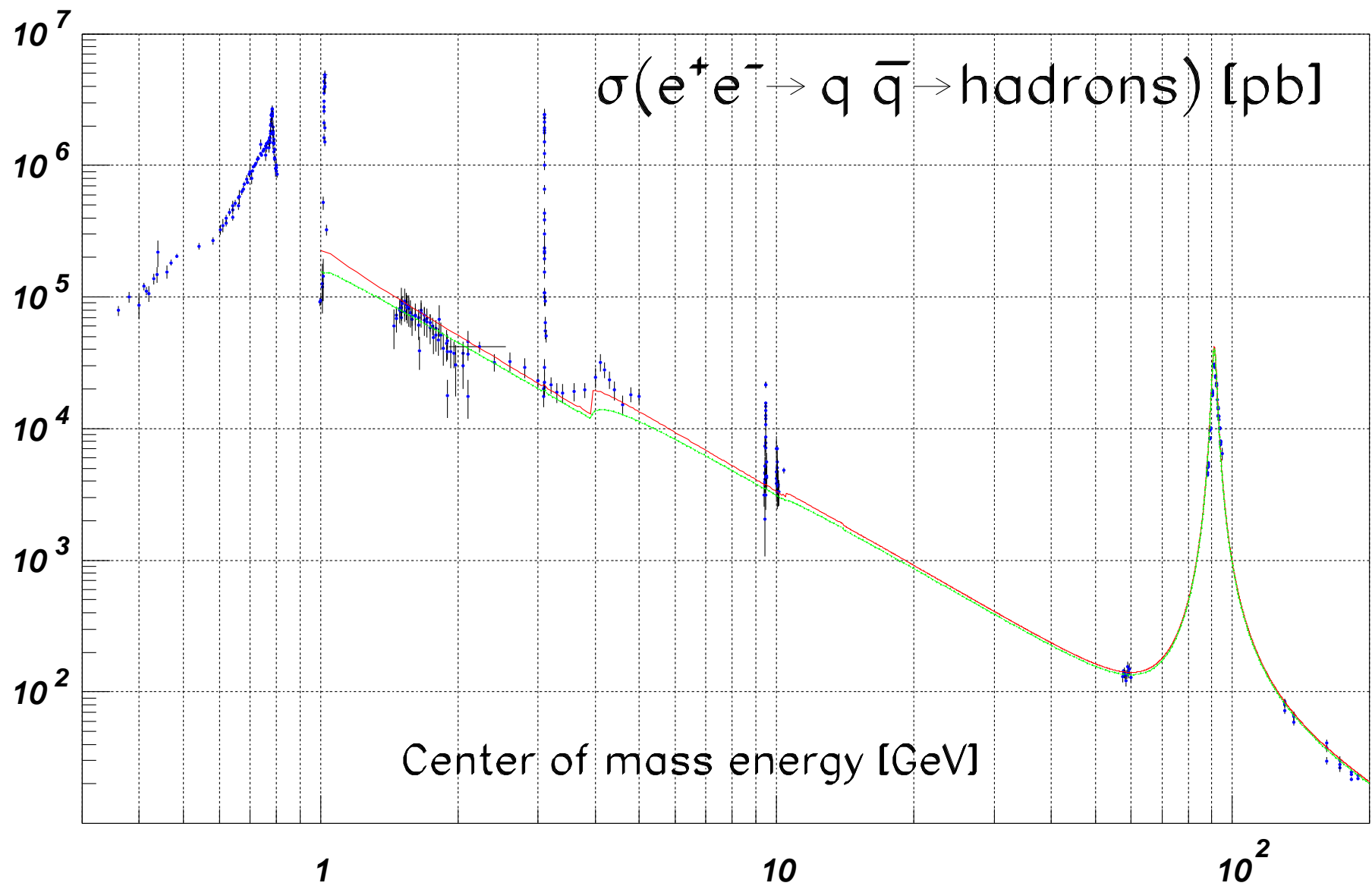
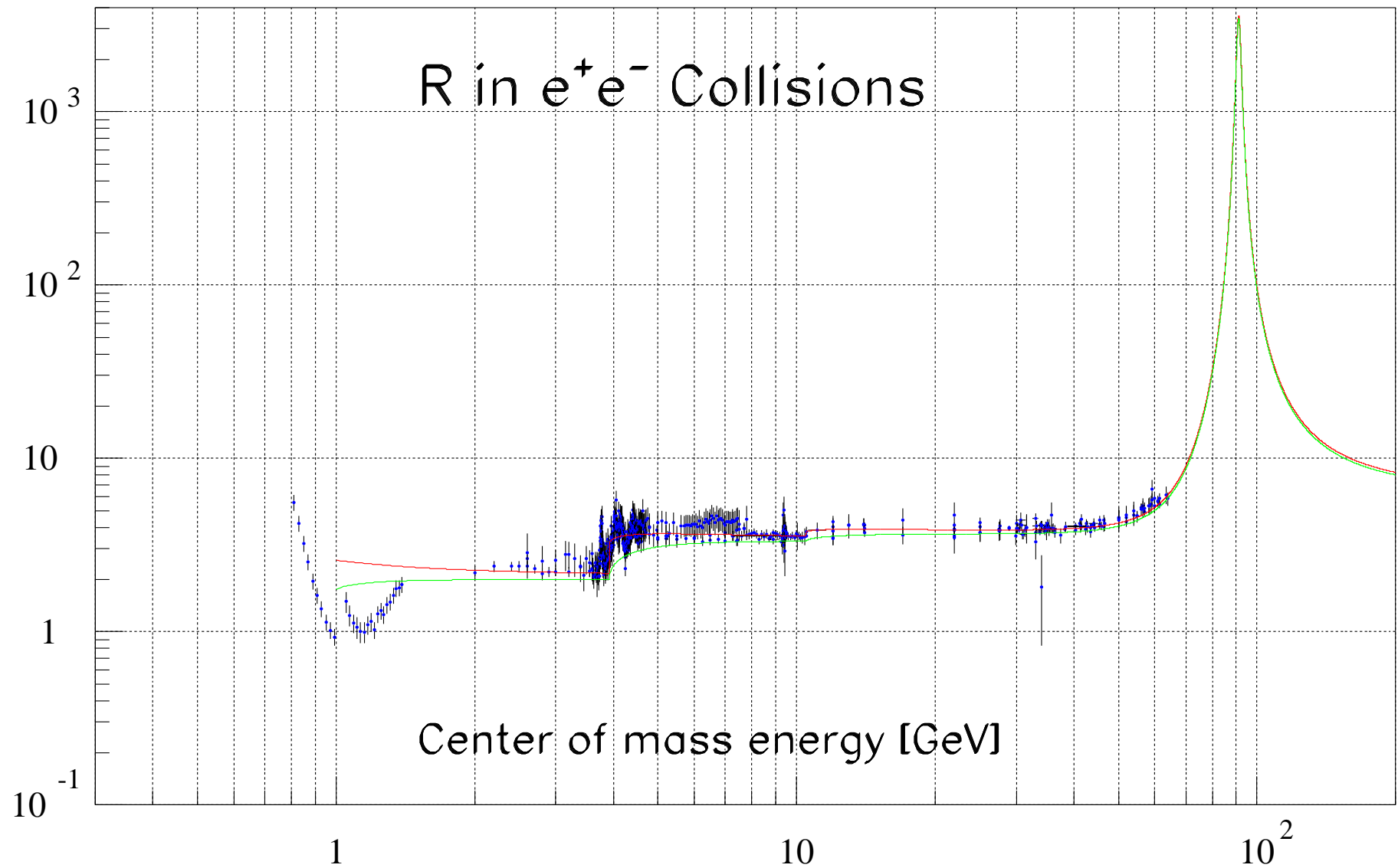
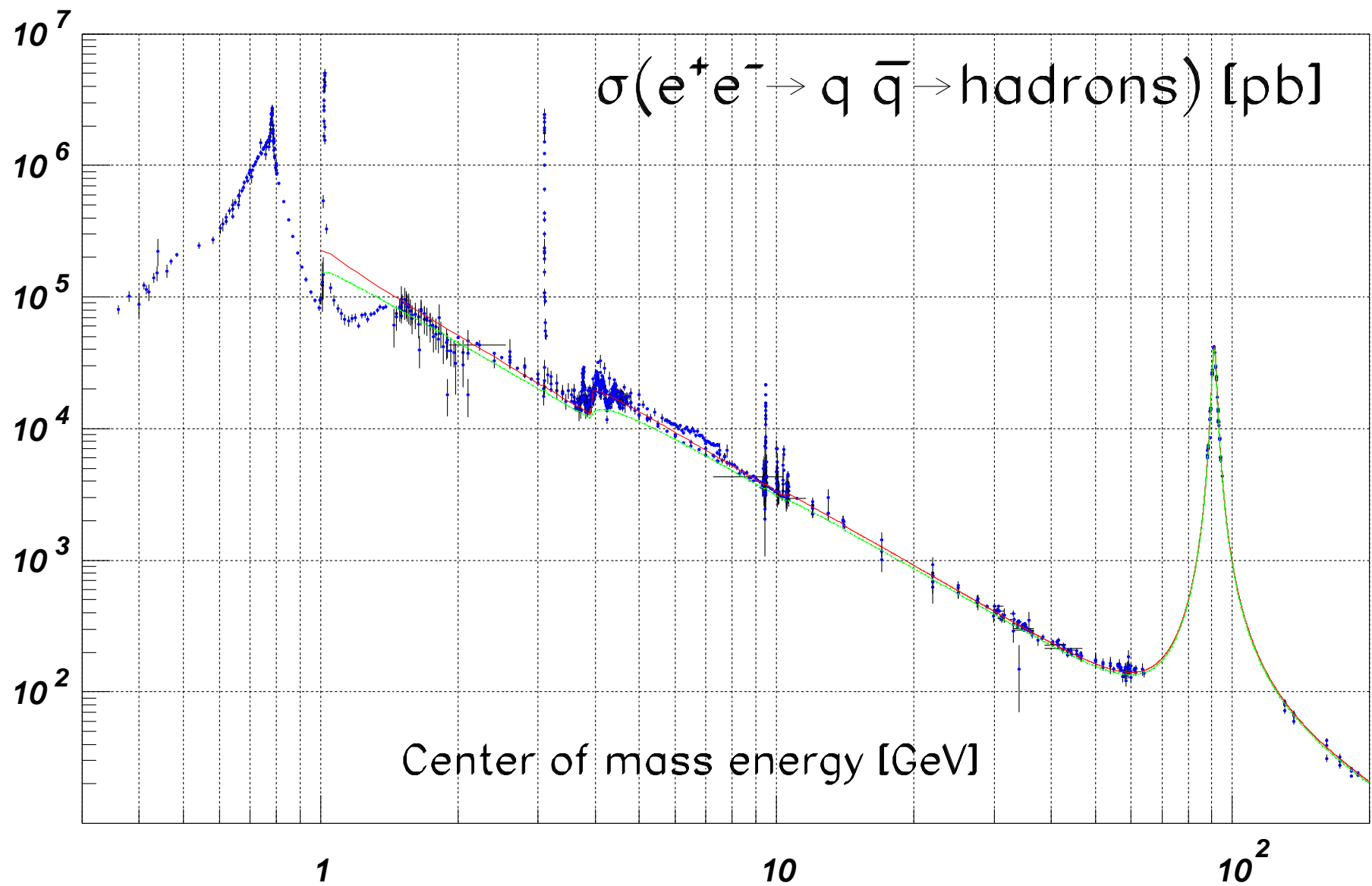
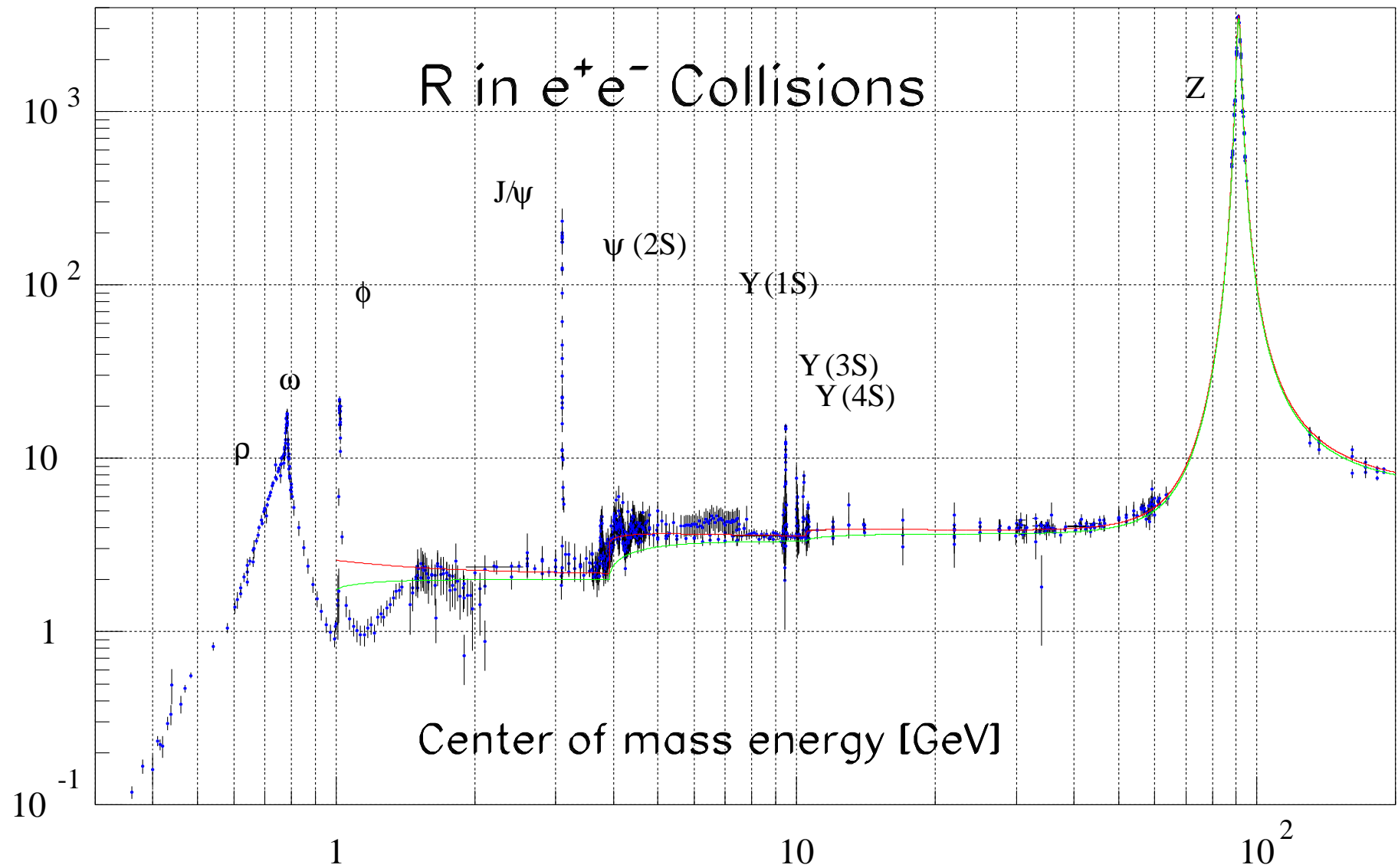
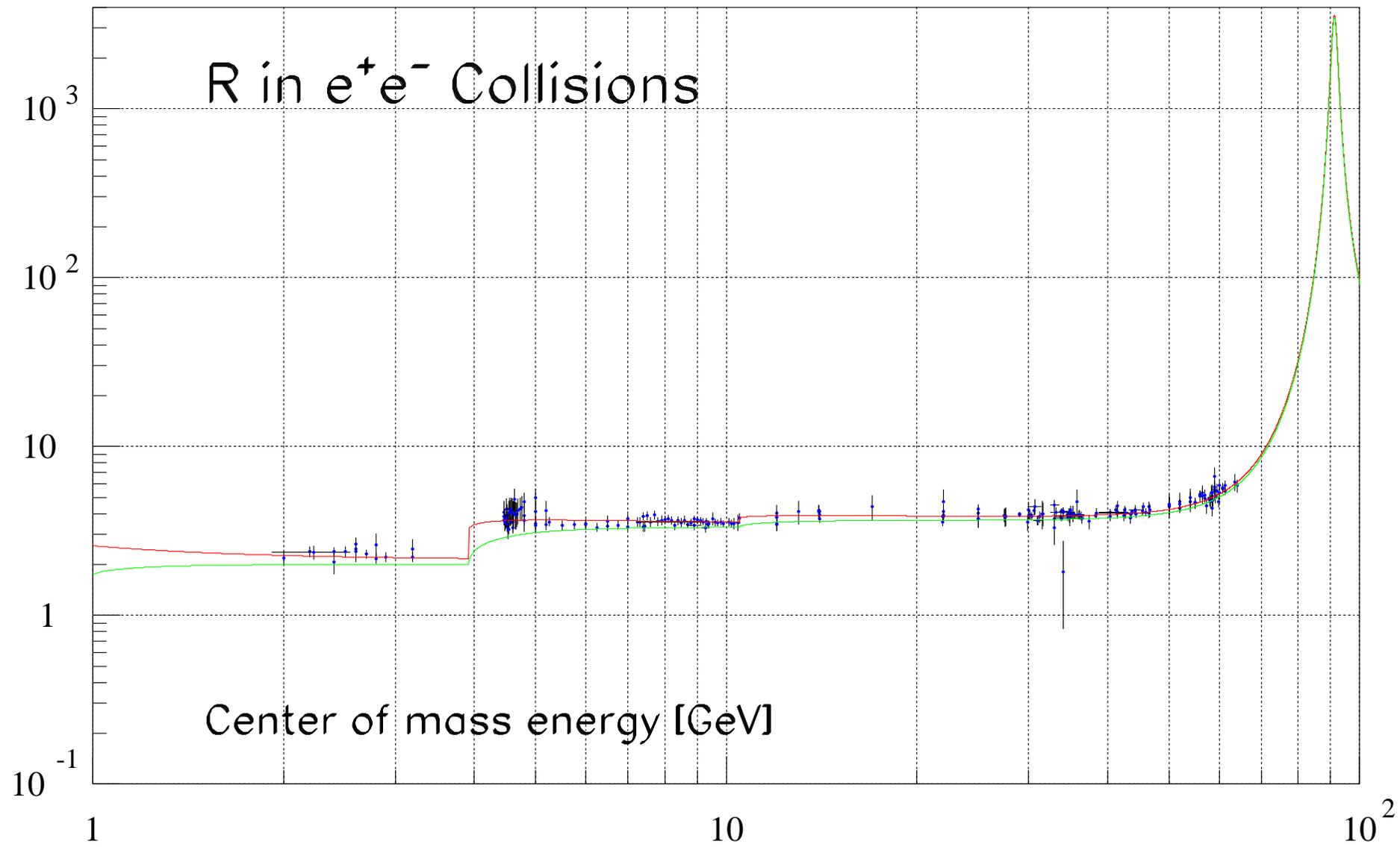
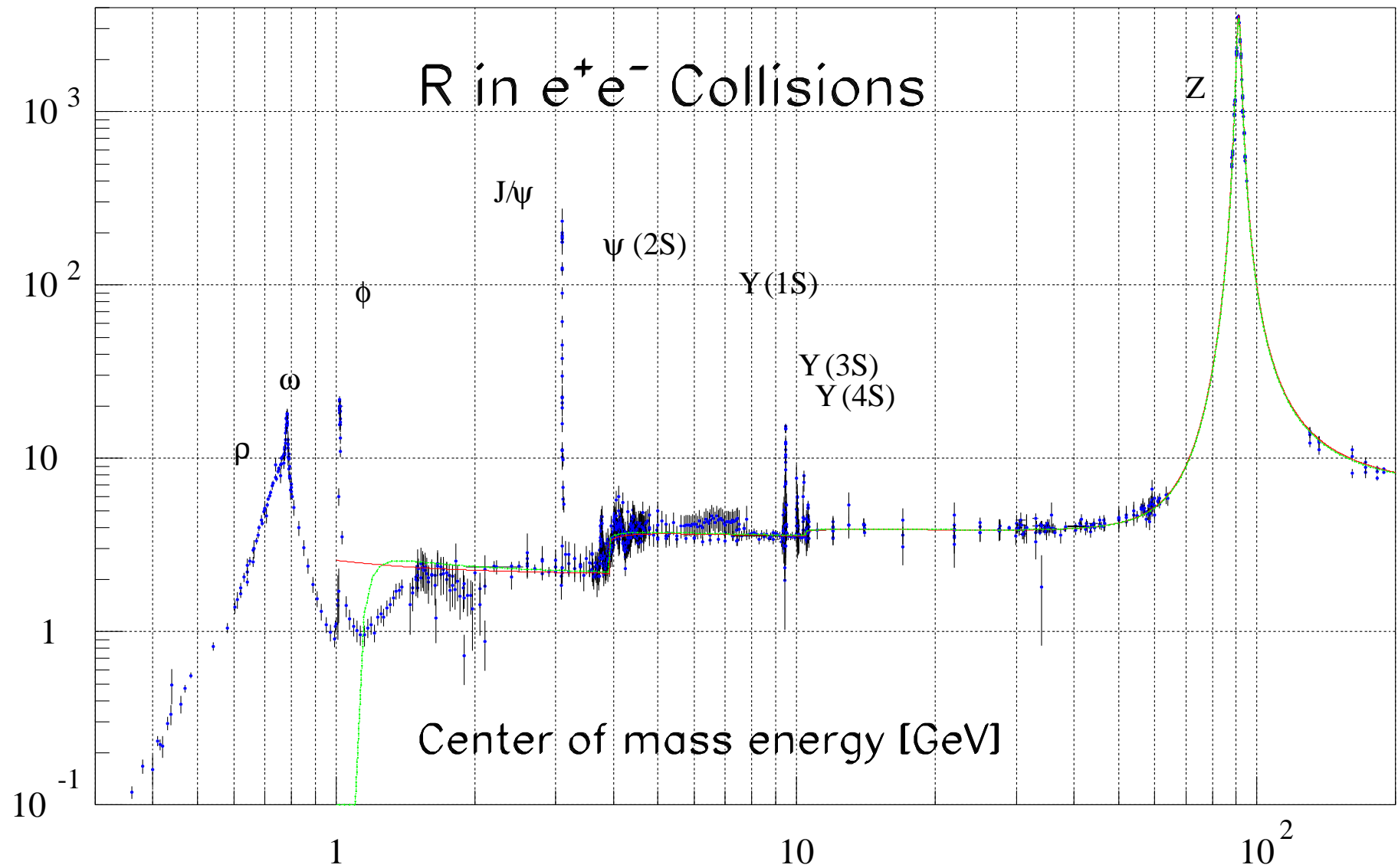
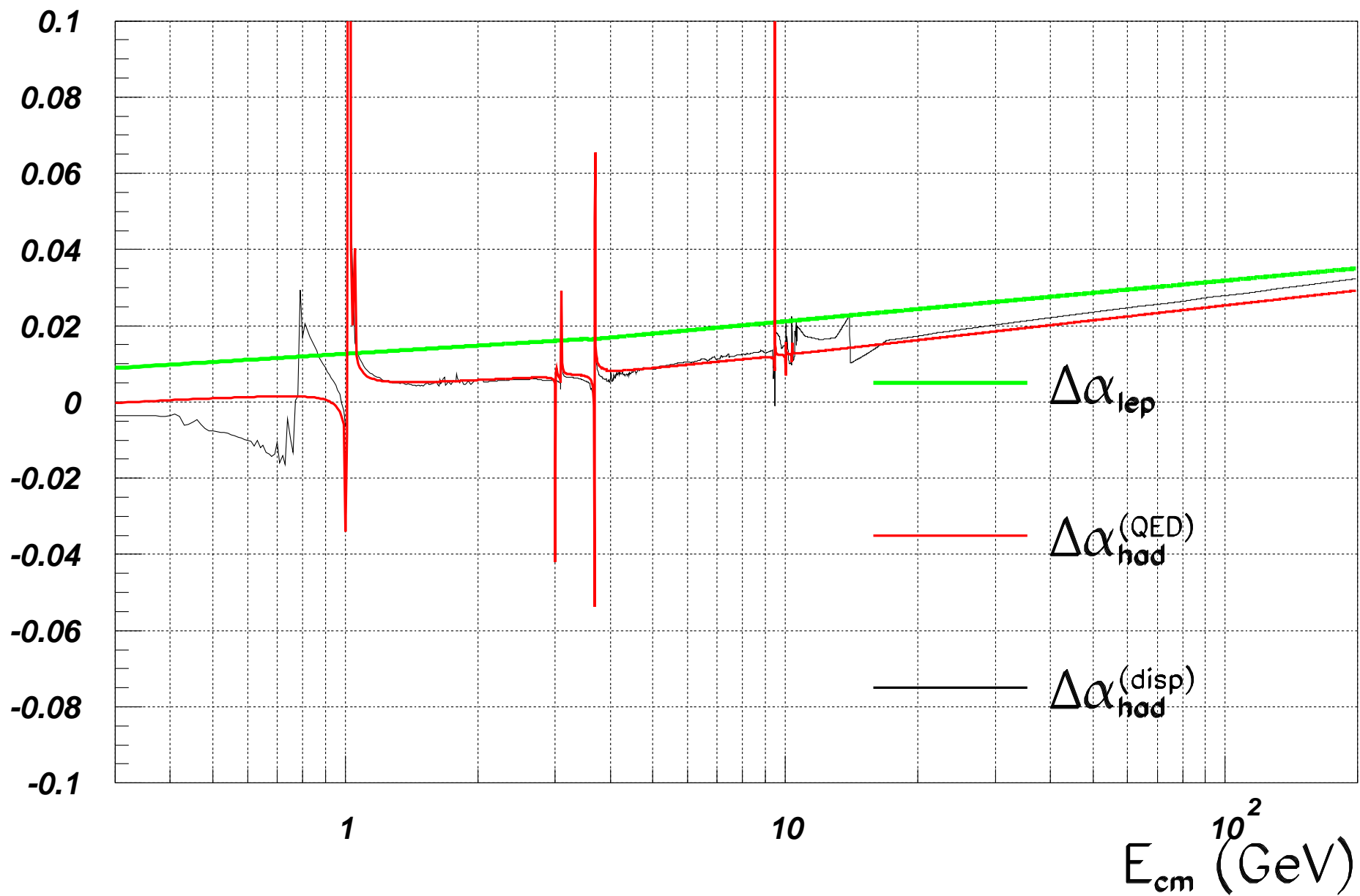
\end{document}